\documentclass[12pt]{article}
\usepackage{amsmath}
\usepackage{amsfonts}
\usepackage{amssymb}

\def\rf#1{(\ref{eq:#1})}
\def\lab#1{\label{eq:#1}}
\begin{document}

\begin{center}

\large{\bf A symmetry reduction technique for higher order 
Painlev\'e systems.}
\end{center}
\normalsize
\vskip .4in

\begin{center}
H. Aratyn

\par \vskip .1in \noindent
Department of Physics \\
University of Illinois at Chicago\\
845 W. Taylor St.\\
Chicago, Illinois 60607-7059\\
\par \vskip .3in

\end{center}

\begin{center}
J.F. Gomes and A.H. Zimerman

\par \vskip .1in \noindent
Instituto de F\'{\i}sica Te\'{o}rica-UNESP\\
Rua Dr Bento Teobaldo Ferraz 271, Bloco II,\\
01140-070 S\~{a}o Paulo, Brazil
\par \vskip .3in

\end{center}
\vskip .2in \noindent
\begin{center}
{\large {\bf ABSTRACT}}\\
\end{center}
\par \vskip .15in \noindent

The symmetry reduction of higher order Painlev\'e systems
is formulated in terms of Dirac procedure.
A set of canonical variables that admit Dirac reduction 
procedure is proposed for Hamiltonian structures  governing the ${A^{(1)}_{2M}}$ 
and ${A^{(1)}_{2M-1}}$  Painlev\'e systems 
for $M=2,3,{\ldots} $.

\newpage

\subsection{Introduction}
Although first introduced in mathematics for their special singularity
properties, Painlev\'e equations are now ubiquitous in physics.
The list of their physical applications
includes such topics as: the Ising and antiferromagnet models  \cite{ising}, 
 the transport of particles across boundaries 
(Nernst-Planck equations) \cite{nernst}, dilute Bose-Einstein condensates in an 
external one dimensional field (Gross-Pitaevskii equation)
\cite{bose-einstein},  Hele-Shaw problems in viscous fluids 
\cite{fokas} as well as various approaches to 
quantum field theory and topological field theory, 
supersymmetric gauge theories, random matrix theory, 
statistical mechanics, plasma
physics, superconductivity, nonlinear optics and fiber optics,  resonant oscillations in 
shallow water, and polymers. 
The $A^{(1)}_3$ Painlev\'e system, corresponding to 
the Painlev\'e V equation, appears in the context
of the reduced density
matrix of the impenetrable Bose gas model \cite{jimbo} and in connection of
a unitary matrix model formulation of chiral Yang-Mills theory on the sphere 
\cite{szabo}.
Also, the Painlev\'e  equations emerge as self-similarity reductions of 
well-known soliton equations. For instance, the 
Painleve system of $A^{(1)}_2$ type, 
corresponding to the Painlev\'e IV equation, can be obtained 
by reduction of the so-called AKNS soliton model.

In this note we will deal with generalization of the Painlev\'e equations of $A^{(1)}_2$ and   $A^{(1)}_3$ types, mentioned above, to the higher order  ${A^{(1)}_{n}}$ type of Painlev\'e system  
of ordinary differential equations 
for $n + 1$ functions $f_0 , . . . , f_n$  and variables $\alpha_0, ..., \alpha_n$
for $n=2,3, {\ldots}$.
An affine Weyl group  ${A^{(1)}_{n}}$ 
acts on a parameter space of system's variables  \cite{NoumiYamada98a,NoumiYamada98b,Noumiwkb,noumibk,sasano}.

The higher order Painlev\'e equations are either quadratic or cubic 
depending on whether $n$ is even or odd.
For $n=2M$ ($M=1,2,\ldots$) equations are given by
\begin{equation}\lab{A2M}
A^{(1)}_{2M}:\qquad f_{i,\, x}=f_i\big (\sum_{r=1}^{M}f_{i+2r-1}
-\sum_{r=1}^{M} f_{i+2r}\big)+\alpha_i\, ,
\end{equation}
where $0\le i \le 2M$ and conditions $f_0 +\cdots+f_n=-2x$ and
$\alpha_0+\cdots+ \alpha_n=-2$ hold for system's functions and parameters.
For $n=2M-1$ ($M=2,\ldots$) the system is described by:
\begin{align}
A^{(1)}_{2M-1}:\quad 2 x f_{i,\,x} &=f_i\big(
\sum_{ 1\le r\le s\le M-1} f_{i+2r}f_{i+2s+1}
-\sum_{ 1\le r\le s\le M-1 } f_{i+2r-1}f_{i+2s}
\big) \lab{A2M+1}
\\ 
&\quad+(-1)^{i+1} f_i \,\big(\sum_{r=0}^{M-1}\alpha_{2r+1}+2\big)  -
\alpha_i\big(\sum_{r=0}^{M-1}f_{i+2r}\big),
\nonumber
\end{align}
with $0\le i\le 2M-1$ and with conditions $\sum_{r=0}^{M-1} f_{2r}=-2x$,
$\sum_{r=0}^{M-1} f_{1+2r}=-2x$ ,  $\sum_{r=0}^{2M-1} \alpha_{r}
=-4$ for system's functions and parameters. 
These conditions differ from that used in  \cite{sasano}   by different normalization of  system's functions. 

The above symmetric Painlev\'e equations are invariant under
the following actions :
\begin{xalignat}{2}
s_i(\alpha_j)&=\alpha_j-a_{ji} \alpha_i ,& \quad 
s_i(f_{i\pm 1})&= f_{i\pm 1}\pm \frac{\alpha_i}{f_i} \lab{affWeyl}\\
\pi (f_j) &= f_{j+1} ,& \quad \pi (\alpha_j)&= \alpha_{j+1} ,
\nonumber
\end{xalignat}
of the extended affine Weyl group ${A^{(1)}_{n}}$ generators $\pi, s_i,
i=0,1,{\ldots}, n$, where $a_{ij} $ are coefficients
of the $A^{(1)}_n$ Cartan matrix $A=(a_{ij})_{0\le i,j\le n}$
\cite{NoumiYamada98a,NoumiYamada98b,Noumiwkb,noumibk}.

The  Painlev\'e equations \rf{A2M} can be realized through a polynomial Hamiltonian 
system  \cite{NoumiYamada98b,sasano} in functions $f_i$   
with the underlying Poisson brackets  :
\begin{equation}
\left\{ f_i\, , \, f_{i+1} \right\} = 1, \qquad
\left\{ f_i\, , \, f_{i-1} \right\} =  -1, \;\; i=1,{\ldots} ,2M \,.
\lab{fifj}
\end{equation}
It is convenient to express the relevant Hamiltonian
formalism in terms of canonical coordinates defined through relations
\begin{equation}
p_i = f_{2i}, \qquad q_i= \sum_{k=1}^i f_{2k-1},\qquad  i=1,{\ldots}
,2M \, ,
\lab{pqf}
\end{equation}
that map Poisson brackets \rf{fifj}  into 
the  canonical brackets  $\{q_i,p_j\}=\delta_{ij}$.
In terms of canonical coordinates $p_i, q_i$ the 
$A^{(1)}_{2M}$ Hamiltonian is given by \cite{NoumiYamada98b,sasano,AGZ:2011} :
\begin{equation} 
\begin{split}
{\cal H}_{A^{(1)}_{2M}}&= \sum_{j=1}^M p_j q_j  \left( p_j+q_j +2 x\right)
+2 \sum_{1\le j <i \le M} p_j q_j p_i\\
&-\sum_{j=1}^M {\alpha}_{2j} q_j + \sum_{j=1}^M p_j \left(\sum_{k=1}^j
\alpha_{2k-1}\right)
\end{split}
\lab{calhmpq}
\end{equation} 
The corresponding Hamilton equations:
\begin{equation} 
\begin{split}
q_{i, \,x} &= \frac{\partial {\cal H}_M}{\partial p_i}= q_i  \left( q_i+2 \sum_{j=i}^M 
p_j +2 x\right) +2 \sum_{j=1}^{i-1} p_jq_j+\sum_{j=1}^i \alpha_{2j-1} \\
p_{i, \,x} &= -\frac{\partial {\cal H}_M}{\partial q_i}= - p_i  \left( 2q_i+p_i+2
\sum_{j>i}^M p_j +2 x\right) + \alpha_{2i}
\lab{hameqs}
\end{split}
\end{equation} 
are mapped through relations \rf{pqf}
into the $A^{(1)}_{2M}$ Painlev\'e equations \rf{A2M}.


In reference \cite{AGZ:2011} we considered a specific integrable model, namely
the 2M-Bose model from constrained KP hierarchy, to construct the
$A_{2M}^{(1)}$ symmetric Painlev\'e equations.  
We studied their symmetries
from the Lax point of view and obtained the corresponding
B\"{a}cklund transformations.
In this paper by defining special combinations of 
coefficients  of Lax operators from  \cite{AGZ:2011} as canonical variables we are able to
show that Dirac  reduction governs the symmetry reduction $A^{(1)}_{2M} \to 
A^{(1)}_{2M-1}$ of the respective Painlev\'e system by reducing   
the Hamiltonian ${\cal H}_{A^{(1)}_{2M}}$ from the definition
\rf{calhmpq} to the Hamiltonian ${{\cal H}}_{A^{(1)}_{2M-1}}$
\cite{NoumiYamada98b,sasano}
given by relation
\begin{align}
2 x { {\cal H}}_{A^{(1)}_{2M-1}}  &= \sum_{j=1}^{M-1}p_j 
\left(p_j+2x\right)
q_j \left(q_j+2x\right)+2\sum_{1\le j <i \le M-1} p_jq_jp_i
\left(q_i+2x\right) \nonumber \\&+
\left(\sum_{k=0}^{M-1} \alpha_{2k+1}+2\right) 
\left(\sum_{j=1}^{M-1} p_jq_j\right) \lab{hrmpq}
\\&- \sum_{j=1}^{M-1} \alpha_{2j} 2 x q_j +\sum_{j=1}^{M-1}
\left(\sum_{k=1}^j \alpha_{2k-1}\right) 2 x p_j
\nonumber
\end{align}
with relations $p_i = f_{2i},\, q_i= \sum_{k=1}^i f_{2k-1}$
for $i=1,{\ldots} ,M-1$ between canonical variables $p_i, q_i,\,
i=1,{\ldots} ,M-1$ of the reduced phase space and 
the Painlev\'e functions $f_1, {\ldots} , f_{2M-2}$ from equation \rf{A2M+1}.
In addition, functions $f_0$ and $f_{2M-1}$ appearing 
in equation \rf{A2M+1} are defined through
$f_0= -2x-f_2-{\ldots} -f_{2M-2}$ and 
$f_{2M-1}=- 2x-f_1-{\ldots} -f_{2M-3}$
resulting in the total of $n+1=(2M-1)+1=2M$ Painlev\'e functions.
Note, that $\alpha_0$ does not enter expression in \rf{hrmpq}
and is determined from the definition $\alpha_0= -4-\alpha_1
-{\ldots} -\alpha_{2M-1}$.

\subsection{Symplectic Map and New Canonical Variables}
Before we present a set of new canonical variables 
to describe the higher order Painlev\'e system let us comment 
on a possibility of applying Dirac reduction procedure
in the framework of canonical variables $p_i,q_i,i=1,..,2M$ of
the ${\cal H}_{A^{(1)}_{2M}}$ Hamiltonian 
given in \rf{calhmpq}. 
A straightforward attempt that involves setting one of the canonical 
variables to zero 
inevitably leads to the dimension of the underlying phase space 
getting reduced by two degrees.
For example, imposing a constraint $p_M=0$ leads 
via the secondary constraint \cite{claudio} $p_{M,\,x}=0$  
to $\alpha_{2M}=0$ as seen from equation \rf{hameqs}. 
That in turn eliminates a presence of $q_M$ from the
Hamiltonian  ${\cal H}_{A^{(1)}_{2M}}$ and effectively causes a reduction 
${\cal H}_{A^{(1)}_{2M}} \to {\cal H}_{A^{(1)}_{2M-2}}$.

Below we describe a different reduction scheme 
leading to the advertised reduction 
${\cal H}_{A^{(1)}_{2M}} \to {\cal H}_{A^{(1)}_{2M-1}}$
It employs a set of  new canonical variables
$e_i,Y_i, i=1,{\ldots} ,M$ obtained  from $p_i,q_i$ through the
the following definitions :
\begin{equation}
\begin{split}
e_M &= p_M+q_M+2x, \qquad
e_{M-1} = -q_{M-1}\\
e_{M-2k}&=-p_k-p_{k+1}-\cdots -p_{M-k-1}\\ 
e_{M-2k-1}&=-q_{M-k-1}+q_k, \quad k=1,2,..
\end{split}
\lab{eMMkr}
\end{equation}
and
\begin{equation}
\begin{split}
Y_M &=  p_M+q_{M-1}+2x, \qquad
Y_{M-1} =-q_{M}-p_M-p_{M-1}\\
Y_{M-2k} &= -q_{k-1}+q_{M-k-1}+2x \\
Y_{M-2k-1} &=P_{k}^M-P_{M-k-1}^M+2x,
\quad k=1,2,3,{\ldots} 
\end{split}
\lab{cMMkr}
\end{equation}
where to simplify expressions we introduced auxiliary quantities :
\begin{equation}
 P^M_k\equiv p_k +p_{k+1}+{\ldots} +p_M, \quad P^M_{M-k}\equiv p_{M-k} +p_{M-k+1}+{\ldots} +p_M,
\lab{qmk}
\end{equation}
inter-connected via relation
\begin{equation}
 P^M_k=P^M_{M-k} -e_{M-2k}\, .
\lab{ppmk}
\end{equation}
The above map, suggested by the study \cite{AGZ:2011} of 
Painlev\'e systems in the context of integrable Voletrra-type
lattices, is symplectic. The canonical
brackets  $\{q_i,p_j\}=\delta_{ij}$ are being transformed
into the usual canonical 
Poisson brackets for $e_i, Y_i$ :
\begin{equation}
\left\{ e_i \, ,\, Y_j \right\}= \delta_{ij}, \;\;\;
\left\{ e_i\, , \, e_{j} \right\} = 0=
\left\{ Y_i\, , \, Y_{j} \right\}, \;\;\;\; i,j=1,{\ldots} ,M
\lab{eYbracket}
\end{equation}
To rewrite the  ${\cal H}_{A^{(1)}_{2M}}$ Hamiltonian 
in \rf{calhmpq} in terms of these new variables it is convenient 
to use inverse relations that give $p_i,q_i$ in terms of $e_i, Y_i$.
The variables $q_k$ are related to new
variables as follows.
\begin{equation}
q_k=-2kx+\sum_{i=1}^k Y_{M-2i} + \sum_{i=2}^{2k+1} e_{M-i},\quad
k=1,2,{\ldots} , \left[M/2\right]-1\, ,
\lab{mm2}
\end{equation}
where $ \left[M/2\right] = M/2$ or $ \left[M/2\right] =(M -1)/2$, 
whichever is an integer. Furthermore,
\begin{equation}
q_{M-k-1}=-2kx+\sum_{i=1}^k Y_{M-2i}+ \sum_{i=2}^{2k} e_{M-i},
\lab{mm21}
\end{equation}
for $k=1,2,{\ldots} , \left[M/2\right]-1$ for even $M$ and $k=1,2,{\ldots} , 
\left[M/2\right]$
for odd $M$.
Finally, 
\begin{equation}
q_M=-Y_M-e_{M-1}, \quad q_{M-1}=-e_{M-1}.
\lab{mmm2}
\end{equation}
Thus the relation to new variables becomes
\begin{equation}
P^M_{k} =  -q_{k-1}+\sum_{i=0}^{2k-1} (-1)^i
\left(Y_{M-i}-e_{M-i}\right),\quad
k=1,2,{\ldots} , \left[M/2\right]-1
\lab{mpp2}
\end{equation}
The relations  for remaining indices are then obtained  through 
\rf{ppmk} :
\begin{equation}
p_M=e_{M-1}+e_M+Y_M-2x, \quad p_{M-1}=-e_M-Y_{M-1}+2x\, .
\lab{mmp2}
\end{equation}
{}For a specific example of $M=5$ the above relations give  :
\begin{alignat}{3}
q_5&= -Y_5 -e_4 , &\qquad p_5&= e_4+e_5+Y_5-2x \nonumber\\
q_4&= -e_4 , &\qquad p_4&= -e_5-Y_4+2x \nonumber\\
q_3&= Y_3 -2x , &\qquad p_3&= -e_3-Y_2+2x \nonumber\\
q_2&= Y_1+Y_3+e_2 -4x , &\qquad p_2&= -e_1 \nonumber\\
q_1&= e_2+Y_3 -2x , &\qquad p_1&= e_1+Y_2-2x \nonumber
\end{alignat}
and it is easy to explicitly verify a symplectic nature of the above map.

In terms of $M$ variables $e_i, Y_i$  the Hamiltonian 
${\cal H}_{A^{(1)}_{2M}}$ becomes :
\begin{align}
{\cal H}_{A^{(1)}_{2M}} &=- \sum_{j=1}^M e_j  \left(Y_j -2x \right) \left(Y_j -e_j\right)
-2\sum_{1\le j <i \le M} (-1)^{i+j}  e_j\left(Y_j -2x \right)
\left(Y_i-e_i\right)\nonumber\\
&+\sum_{j=1}^M {\bar k}_j Y_j 
- \sum_{j=1}^M    {\kappa}_j  e_j
\lab{hmeY}
\end{align}
with constants ${\kappa}_j, {\bar k}_j, j=1,{\ldots} ,M$ related to
Painlev\'e parameters via:
\begin{equation}
\begin{split}
\kappa_j&=  \underbrace{\alpha_{M-j-1}+ \alpha_{M-j+1}+\cdots
+ \alpha_{M+j-3}}_{j} , \quad j=1,{\ldots} , M-2 \\
\kappa_{M-1}&=-\sum_{j=1}^{M} \alpha_{2j-1} -\alpha_{2M-2}-\alpha_{2M}
, \qquad 
\kappa_{M}=\sum_{j=1}^{M-1} \alpha_{2j-1} +\alpha_{2M}\\
{\bar k}_l &= \underbrace{-\alpha_{M-l}- \alpha_{M-l+2}- \cdots
- \alpha_{M+l-2}}_{l} , \quad l=1,{\ldots} , M-1,\\\
{\bar k}_M &= \sum_{j=1}^{M} \alpha_{2j-1}+\alpha_{2M}
\end{split}
\lab{consts-unred}
\end{equation}

\subsection{From $A^{(1)}_{2M}$ to $A^{(1)}_{2M-1}$ via Dirac Reduction}
We proceed by imposing a constraint:
\begin{equation} 
Y_M=0\, .
\lab{ym0}
\end{equation}
The consistency requires that we also need to impose the secondary
constraint (see e.g.  \cite{claudio}) :
\begin{equation} 
0=Y_{M,\,x} = - \frac{\partial }{\partial e_M} {\cal H}_M= 
4xe_M -2 \sum_{i=1}^{M-1} (-1)^{M-i}e_i \left( Y_i-2x \right) 
+\kappa_M 
\lab{ymx0}
\end{equation}
or
\begin{equation} 
e_M=\frac{1}{4x} \left( 2 \sum_{i=1}^{M-1} (-1)^{M-i}e_i 
\left( Y_i-2x \right) 
-\kappa_M \right) \, .
\lab{emcond}
\end{equation}
Substituting the value of $e_M$ from eq. \rf{emcond} and $Y_M=0$ 
into ${\cal H}_M$ reduces it into 
\[ {\bar {\cal H}}_M = {\cal H}_M \big\vert_{Y_M =0, Y_{M,\,x}=0} \] 
given by  
\begin{equation} 
\begin{split}
2 x {\bar {\cal H}}_M  &= \sum_{j=1}^{M-1} e_j  \left(Y_j -2x \right)
Y_j \left( e_j-2 x \right)\\
&+2\sum_{1\le i< j \le M-1} (-1)^{i+j}  e_i\left(Y_i -2x
\right)Y_j\left(e_j-2x \right)\\
&- {\kappa}_M\sum_{i=1}^{M-1}
(-1)^{M+i} e_i Y_i
+\sum_{j=1}^{M-1} {\bar k}_j 2 x Y_j \\
&- \sum_{j=1}^{M-1}  \left( {\kappa}_j- (-1)^{M+j}
 {\kappa}_M\right) 2 x e_j
\end{split}
\lab{hrmeY}
\end{equation}
Next, we proceed by inserting into the above Hamiltonian ${\bar {\cal H}}_M$ expressions 
\begin{equation}
Y_{M-1}=-p_{M-1},\qquad  e_{M-1} = -q_{M-1}
\lab{Ym1}
\end{equation}
together with relations following from eqs. \rf{eMMkr}-\rf{cMMkr}
for $e_{i}, Y_{i}$ for the remaining indices $i=1,{\ldots} , M-2$ :
\begin{equation}
\begin{split}
e_{M-2k}&=-p_k-p_{k+1}-\cdots -p_{M-k-1}=-P_k^M+P_{M-k}^M\\
e_{M-2k-1}&=-q_{M-k-1}+q_k, \quad k=1,2,..\\
Y_{M-2k} &= -q_{k-1}+q_{M-k-1}+2x \\
Y_{M-2k-1} &=P_{k}^M-P_{M-k-1}^M+2x \, .
\quad k=1,2,3,{\ldots} \lab{cMM2r}
\end{split}
\end{equation}
This casts ${\bar {\cal H}}_M$ into ${\cal H}_{A^{(1)}_{2M-1}}$
from equation \rf{hrmpq} provided that constants
$\kappa_j$ for $j=1,{\ldots} , M-2$ and ${\bar k}_l$
for $l=1,{\ldots} , M-1$ agree with values given in relations
\rf{consts-unred} and in addition relations $\kappa_{M-1}
=-\sum_{j=1}^{M} \alpha_{2j-1}-\alpha_{2M-2} -2$
and $\kappa_{M}=\sum_{j=1}^{M} \alpha_{2j-1} +2$ hold as well.

\subsection{Example of reduction for $M=2$. The Painlev\'e V Equation}
We now study in detail an example of $M=2$.
Although this model and its reduction appeared in  \cite{AGZ:2011},  here we employ new canonical variables to
compactly rewrite the relevant Hamiltonian as
\[
{\cal H}_{A^{(1)}_{4}} =- \sum_{j=1}^2 e_j  \left(Y_j -2x \right) \left(Y_j -e_j\right)
+2  e_1\left(Y_1 -2x \right) \left(Y_2-e_2\right) 
+\sum_{j=1}^2 {\bar k}_j Y_j 
- \sum_{j=1}^2    {\kappa}_j  e_j \,
\]
and to directly implement reduction by substituting $Y_2=0$, 
$e_2=- \left(2 e_1 (Y_1-2x)-\kappa_2\right)/4x$
to obtain the reduced Hamiltonian :
\begin{equation} 
2 x {\bar {\cal H}}_2  =  e_1  \left(Y_1 -2x \right)
Y_1 \left( e_1-2 x \right)
+ {\kappa}_2 e_1 Y_1
+ {\bar k}_1 2 x Y_1 \\
-   \left( {\kappa}_1+{\kappa}_2 \right) 2 x e_1
\lab{hr2eY}
\end{equation}
The corresponding Hamilton equations are :
\begin{equation}
\begin{split}
e_{1\,x} &= 2 x e_1-e^{2}_{1}
-2 e_{1} Y_1 +{\bar k}_1
+\frac{1}{2x} \left( 2e^{2}_{1}Y_1  +e_1 \kappa_2\right)\\
Y_{1\,x} &= -2x Y_1 +2 e_1Y_1+Y_1^2 + \kappa_1+\kappa_2
-\frac{1}{2x} \left( 2Y_1^2 e_1+ Y_1 \kappa_2\right) \, .
\lab{e1Y}
\end{split}
\end{equation}
The above equations are  fully symmetric and invariant under
the B\"acklund transformations:
\begin{equation}
\begin{split}
e_1 & \stackrel{g}{\longrightarrow} Y_1 \\
Y_1 & \stackrel{g}{\longrightarrow} e_1 +Y_1-\frac{Y_{1\,x}}{Y_1} 
- \frac{1}{2x} \left(2e_1 Y_1+\kappa_2 \right)\\
&=-e_1+2x-\frac{\kappa_1+\kappa_2}{Y_1} 
\lab{backe1Y}
\end{split}
\end{equation}
together with appropriate transformations of constants given 
in relation 
\begin{equation} \begin{split}
g(\kappa_1) &= 2-\kappa_1-\kappa_2-{\bar k}_1\\
g(\kappa_2) &= 2\kappa_1+\kappa_2\\
g({\bar k}_1) &= -\kappa_1-\kappa_2
\lab{gks}
\end{split}
\end{equation}
{}From eqs. \rf{Ym1} it follows that  $Y_1=-p_1=-f_2$ and 
$e_1=-q_1=- f_1$. Subsequently $f_0=-2x+Y_1$ and
$f_3=-2x+e_1$. In this notation one can cast equations of motion \rf{e1Y}
into symmetric $A^{(1)}_3$ Painlev\'e V equations:
\begin{equation}
\begin{split}
2x f_{0\,x}&=f_0 f_2 \left(f_3-f_1\right) -\left(\alpha_1+\alpha_3+2\right) 
f_0 - \alpha_0 (f_0+f_2)\\
2x f_{2\,x}&=f_0 f_2 \left(f_1-f_3\right) -\left(\alpha_1+\alpha_3+2\right) 
f_2 - \alpha_2 (f_0+f_2)\\
2x f_{1\,x}&=f_1 f_3 \left(f_0-f_2\right) +\left(\alpha_1+\alpha_3+2\right) 
f_1 - \alpha_1 (f_1+f_3)\\
2x f_{3\,x}&=f_1 f_3 \left(f_2-f_0\right) +
\left(\alpha_1+\alpha_3+2\right) f_3 - \alpha_3 (f_1+f_3)
\lab{nagoya}
\end{split}
\end{equation}
with constants :
\[\alpha_1=-{\bar k}_1,\;\;\; \alpha_2=-\kappa_1-\kappa_2,\;\;\; 
\alpha_3=-2 +\kappa_2+{\bar k}_1
\]
and $\alpha_0 =-4 -\alpha_1-\alpha_2-\alpha_3=-2+\kappa_1$. 
In this setting the 
B\"acklund transformation $g$ defined through relations \rf{gks} and
\rf{backe1Y} is identified with $g=\pi s_1$, where actions
of $\pi$ and $s_1$ are defined in \rf{affWeyl} for $n=3$.
\subsection{Outlook}
We have proposed a set of new canonical variables
for 
Hamiltonian formalism for 
higher Painlev\'e equations that enables symmetry reduction by Dirac
procedure.  It would be interesting to
extend such construction beyond the class of $A^{(1)}_n$ type of equations
treated in this paper and to use the formalism presented here 
to deal with questions
of how to formulate Hamiltonian structures for various symmetry groups
to study possible reduction mechanisms.


\begin{thebibliography}{99}
\bibitem{bose-einstein}
A. Aftalion, Q. Du and Y. Pomeau, 
Phys. Rev. Lett. {\bf 91}, 090407 (2003).
\bibitem{AGZ:2011}
  H.~Aratyn, J.~F.~Gomes and A.~H.~Zimerman,
  J.\ Phys.\  A: Math. Theor. {\bf 44}, 235202 (2011).
\bibitem{nernst}
A J Bracken, L Bass and C Rogers, 
 J. Phys. A: Math. Theor. {\bf 45} 105204 (2012),
 [arXiv:math-ph/1201.0673]
\bibitem{claudio}
A. Hanson, T. Regge and C. Teitelboim, Constrained Hamiltonian Systems
(Accademia Nazionale dei Lincei, Rome, 1976);
M. Henneaux and C. Teitelboim, Quantization of Gauge Systems
(Princeton University Press, Princeton) 1992
\bibitem{fokas}
A.S. Fokas and S. Tanveer,
Math. Proc. Cambridge Philos. Soc., {\bf 124}, no.1, 169-191 (1998)
\bibitem{jimbo} M. Jimbo, T. Miwa, Y. Môri and M. Sato, 
Physica {\bf D 1} (1980) 80-158. 
\bibitem{ising}
B. M. McCoy, C. A. Tracy and T. T. Wu, 
Phys. Rev. Lett. {\bf 38} (1977) 793-796. 
\bibitem{NoumiYamada98a}
M. Noumi and Y. Yamada, 
Comm. Math. Phys. {\bf 199},  281--295 (1998)
\bibitem{NoumiYamada98b}
M. Noumi and Y. Yamada, 
Funkcial. Ekvac. {\bf 41}, 483--503 (1998)
\bibitem{Noumiwkb}
M. Noumi and Y. Yamada, Affine Weyl Group Symmetries in Painlev\'e ́ 
Type Equations, in: C.J. Howls, T. Kawai and Y. Takei (eds.), Toward the 
Exact WKB Analysis of Differential Equations, Linear or Non-Linear, 
pp. 245--259 (Kyoto University Press, Kyoto, 2000);
M. Noumi,
Affine Weyl Group Approach to Painlev\'e Equations,
Proceedings of the ICM, Beijing 2002, vol. 3, 497--510;
arXiv:math-ph/0304042
\bibitem{noumibk}
M. Noumi, Painlev\'e Equations through Symmetry, in: Translations of
Mathematical Monographs, vol. 223, American Mathematical Society
Providence, RI, 2004.
\bibitem{sasano}
Y. Sasano and Y. Yamada, Symmetry and Holomorphy of Painlev\'e Type Systems, RIMS. Kokyuroku 
B{\bf 2}, 215--225 (2007)
\bibitem{szabo}
R. J. Szabo and M. Tierz,
J. Phys. A: Math. Theor. 45 (2012) 085401 [arXiv:hep-th/1102.3640]
\end{thebibliography}
\end{document}